\documentclass[prb,twocolumn,showpacs,superscriptaddress]{revtex4-1}
\usepackage[utf8]{inputenc}
\usepackage[T1]{fontenc}
\usepackage{graphicx}
\usepackage{dcolumn}
\usepackage{bm}
\usepackage{graphicx}
\usepackage{amsmath}
\usepackage{textcomp}
\usepackage{verbatim}
\usepackage{color}
\usepackage{natbib}
\usepackage[symbol]{footmisc}
\usepackage[pdftex,colorlinks=true,linkcolor=blue,urlcolor=blue,citecolor=blue]{hyperref}
\begin{document}
\title{Magnetically controlled exciton transfer in hybrid quantum dot-quantum well nanostructures}
\author{V. Laurindo Jr.}
\affiliation{Departamento de Física, Universidade Federal de São Carlos, 13565-905, São Carlos, São Paulo, Brazil}
\author{Yu. I. Mazur}
\affiliation{Institute for Nanoscience and Engineering, University of Arkansas, Fayetteville, AR 72701, USA}
\author{E. R. Cardozo de Oliveira}
\affiliation{Departamento de Física, Universidade Federal de São Carlos, 13565-905, São Carlos, São Paulo, Brazil}
\author{B. Alén}
\affiliation{Instituto de Micro y Nanotecnología, IMN-CNM, CSIC (CEI UAM+CSIC)Isaac Newton, 8, E-28760, Tres Cantos, 28760, Madrid, Spain}
\author{M. E. Ware}
\affiliation{Institute for Nanoscience and Engineering, University of Arkansas, Fayetteville, AR 72701, USA}
\author{E. Marega Jr.}
\affiliation{Instituto de Física de São Carlos, Universidade de São Paulo, 13.566-590, São Carlos, São Paulo, Brazil}
\author{Z. Ya. Zhuchenko}
\affiliation{Institute of Semiconductor Physics, National Academy of Sciences, pr, Nauki 45, Kiev-03028, Ukraine}
\author{G. G. Tarasov}
\affiliation{Institute of Semiconductor Physics, National Academy of Sciences, pr, Nauki 45, Kiev-03028, Ukraine}
\author{G. E. Marques}
\affiliation{Departamento de Física, Universidade Federal de São Carlos, 13565-905, São Carlos, São Paulo, Brazil}
\author{M. D. Teodoro}
\email[Corresponding author: ]{mdaldin@df.ufscar.br}
\affiliation{Departamento de Física, Universidade Federal de São Carlos, 13565-905, São Carlos, São Paulo, Brazil}
\author{G. J. Salamo}
\affiliation{Institute for Nanoscience and Engineering, University of Arkansas, Fayetteville, AR 72701, USA}


\begin{abstract}
A magnetophotoluminescence study of the carrier transfer with hybrid InAs/GaAs quantum dot(QD)-InGaAs quantum well (QW) structures is carried out where we observe an unsual dependence of the photoluminescence (PL) on the GaAs barrier thickness at strong magnetic field and excitation density. For the case of a thin barrier the QW PL intensity is observed to increase at the expense of a decrease in the QD PL intensity. This is attributed to changes in the interplane carrier dynamics in the QW and the wetting layer (WL) resulting from increasing the magnetic field along with changes in the coupling between QD excited states and exciton states in the QW and the WL.
\end{abstract}

\maketitle

\section{Introduction}
Quantum dot- quantum well heterostructures represent a class of hybrid structures whose photoluminescence wavelength depends on the dot sizes, well width, and the dot-well barrier defining the strength of dot-well coupling. Variation of these parameters provides wide tuneability in the engineering of these systems for optoelectronic applications including high-performance lasers, quantum information processors, or single electron transistors~\cite{1,2,3,4}. For use in high-speed tunnel injection QD lasers, the injected carriers are first collected by the QW, then tunnel into the QDs with subsequent relaxation to the ground state for laser action. By tunneling, cold carriers (electrons) from the QW transfer into the QD states without heating other carries or phonons, thus reducing carrier  leakage from the active region and, hence, increasing the differential gain in the lasers \cite{5, 6, 7}. Optical properties of InAs/GaAs-InGaAs/GaAs dot-well structures have been extensively studied by means of steady-state and time resolved PL, pump-probe measurements clarifying many issues of exciton dynamics \cite{8, 9, 10, 11, 12}. However, application of a magnetic field to such structures allows for additional valuable information to be collected, because it introduces a strong but predictable change to the electronic structure \cite{13, 14}. At high magnetic fields, where the cyclotron energy is larger than both the lateral confinement energy and the exciton binding energy, the magnetic confinement dominates and a Landau-level-like structure is expected to develop. The magnetic field also removes the spin degeneracy giving rise to qualitatively different magnetic field dependences of the emission from the ground and excited states. It is especially important in the case of dot-well structures where excited QD states are brought into resonances with the QW ground or excited exciton states. Moreover, it was revealed recently that the existence of indium-enriched islands in the InGaAs QW results in a spatially indirect (type II) exciton. These excitons electrons bound to a positively charged hole inside a two-dimensional QW move in a ring-like orbit. As a result the PL intensity in InGaAs/GaAs QWs oscillates with magnetic field applied perpendicular to the QW plane at low temperatures. These oscillations have been attributed to the optical Aharonov-Bohn effect associated with spatially indirect excitons that are formed in the vicinity of indium-rich InGaAs islands within the QW\cite{15}.

In this paper we have performed a magnetophotoluminescence (MPL) study of a hybrid InAs/GaAs QD-InGaAs QW structure up to 9 T at a temperature of 4 K. The excitation density was intentionally increased to observe the third QD excited state which is in resonance with the QW exciton ground state. This results in a strong dependence of the luminescence intensities on the magnetic field in the case of thin GaAs barriers.

\section{Sample and Experimental Setup}
A set of hybrid InAs QDs-In$_{0.13}$Ga$_{0.87}$As QW samples were grown on semi-insulating GaAs (001) substrates in a Riber 32 molecular beam epitaxy (MBE) system. Growth details are given in Ref.~\onlinecite{10}. Each sample consists of a 0.3 $\mu$m-thick GaAs buffer layer, a 14-nm-thick InGaAs QW, a GaAs barrier with thickness (d$_{sp}$) of 2 and 20 nm, and a layer of self-assembled InAs QDs covered with 50 nm GaAs cap layer. Two reference samples were grown under the same growth conditions as the dot-well structures. One is the  self-assembled InAs QDs grown on GaAs buffer layer and the other contains only a simple In$_{0.13}$Ga$_{0.87}$As QW. Structural analysis (not shown here) by transmission electron microscopy revealed the InAs QDs to be in the shape of platelets of approximately 5 nm height and 20 nm in diameter on average with an areal density  of $10^{10}$ $cm^2$. The measured thickness of the QW was 14.0 $\pm$ 0.5 nm in all samples and the GaAs spacer thickness was in good agreement with the parameters set during MBE growth.

 	MPL measurements were performed at 4 K and with magnetic fields applied in Faraday geometry with magnitudes up to 9 T using a vibration-free helium closed cycle cryostat (Attocube / Attodry1000) and a home made confocal microscope. A single mode optical fiber with 5 $\mu$m core was used to bring a 660 nm (Toptica / Ibeam Series) with a focus of 1 $\mu$m spot and an excitation power of 90 $\mu$W. The luminescence from the sample was then collected by a multimode 50 $\mu$m optical fiber before being dispersed by a 0.5 m diffractive spectrometer and detected with an InGaAs diode array detector (Andor / Shamrock/Idus). Linear polarizers and half/quarter wave plates were properly set in order to identify the correspondents sigma plus ($\sigma^+$) and minus ($\sigma^-$) optical components emissions from all the samples.

\section{Results and Discussion}
PL spectra measured from the reference QD sample at low temperature (T = 4 K), high excitation power (P= 90 $\mu$W) and different magnetic fields, (B = 0 and 9 T), are shown in Fig. 1(a). At low excitation power, P=12.6 $\mu$W, the PL spectrum of the QDs (not shown) exhibits a single Gaussian emission with a maximum at E =1.135 eV and a full width at half maximum (FWHM) of 34 meV. At high power (P = 90 $\mu$W) the reference QD spectrum transforms into the multiband spectral distribution shown in Fig. 1(a) with the blue line for zero magnetic field. Here the original ground-state is found to red shift by 6 meV due to many-body interaction in the QDs \cite{16}. The additional bands which develop to the high-energy side of the ground-state Gaussian band are assigned to the dipole-allowed interband transition between excited QD states caused by states filling of the lower energy  levels in the QDs. As can be seen from 	Fig. 1(a) in the reference QD sample, a laser power of 90 $\mu$W is enough to fill up to the second excited state at 1.261 eV. Excited state bands are separated by 66 meV and their FWHM ranges from 34 to 47 meV, as deduced from a multiple Gaussian fit. As the magnetic field is increased to 9 T the QD spectra exhibit a blue shift of the ground-state of 1.2 meV, while the line shape and intensity  of the excited states evolve as well. The asymmetry, broadening, and shift of the PL bands can be accurately fitted using multiple gaussians. Such a fit is shown in Fig. 1(a) for $E_{QD}^{0}$ and $E_{QD}^{1}$ transistions in magnetic field B = 9 T for $\sigma^-$ emission. In Ref.~\onlinecite{17} it has been demonstrated that the first excited state can be splitted into two components related to break on the angular momentum degeneracy. Therefore, on this basis the first excited state was deconvoluted into two gaussians and this splitting increases linearly with the magnetic field up to a value of 28.7 meV at B = 9 T. Comparing the spectra measured in $\pm$ configurations, we measure the spin splitting of the QD states arising due to lifting of the spin degeneracy by the magnetic field. This splitting is given by E = $g\mu_{B}B$, where g is the exciton g-factor and $\mu_{B}$ is the Bohr magneton.

\begin{figure}[h!]
\linespread{0.5}
\par
\begin{center}
\includegraphics[scale=0.4]{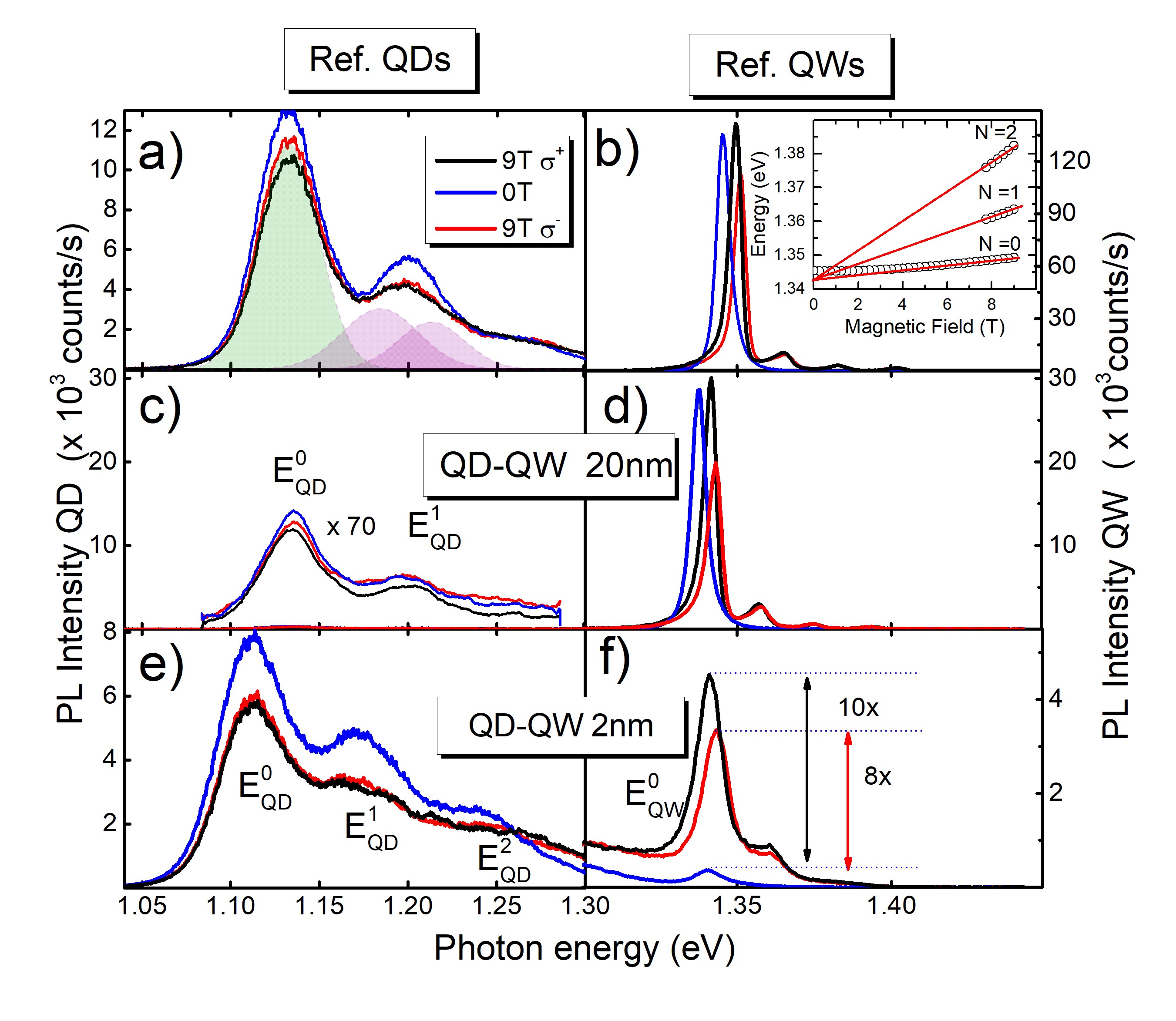}
\end{center}
\par
\vspace{-0.5cm} \caption{ PL spectra measured (a) for reference InAs QDs, (b) reference In$_{0.13}$Ga$_{0.87}$As QW. The inset shows the LLs splitting (solid line are guides to the eye). Hybrid dot-well structure with d$_{sp}$=20 nm (c) QD and (d) QW and  hybrid dot-well structure with d$_{sp}$=2 nm (e) QD (f) QW under high excitation power P$_{ex}$ = 90 $\mu$W. For all datasets $\sigma^+$(black lines) and $\sigma^-$(red lines) represent the detecd polarization at 9 T and without magnetic field (blue lines). Gaussian fit for the ground state and first excited QD states is shown in Fig. 1(a).}
\label{plinten}
\end{figure}

	The low temperature (T = 4 K) PL spectra measured for the reference In$_{0.13}$Ga$_{0.87}$As QW under high excitation power P = 90 $\mu$W for B = 0 T and 9 T are shown in Fig. 1(b). The low excitation PL spectrum (not shown here) has a maximum at E = 1.349 eV and a FWHM value 2.95 meV at zero-field. This is the free exciton transition, $e_{1}-hh_{1}$, in the InGaAs QW. By increasing the magnetic field, the PL band blue-shifts and some asymmetry is introduced. If the excitation intensity increases to P = 90 $\mu$W the PL line shape becomes strongly dependent on magnetic field. High excitation power generates a high areal density of excitons in the QW resulting in Landau quantization at high field as shown in Fig. 1(b). Distinct peaks corresponding to transitions between Landau Levels (LLs) allow us to draw fan-like plots of energies as a function of the magnetic field (inset Fig.1(b)). The total shift of the QW ground state emission reaches 6 meV at 9T, and the spin splitting between $\sigma^{+}$ and $\sigma^{-}$ polarizations is found to be $\sim$ 1.5 meV at $B=9 $T.

	\begin{figure}[h!]
\linespread{0.5}
\par
\begin{center}
\includegraphics[scale=0.4]{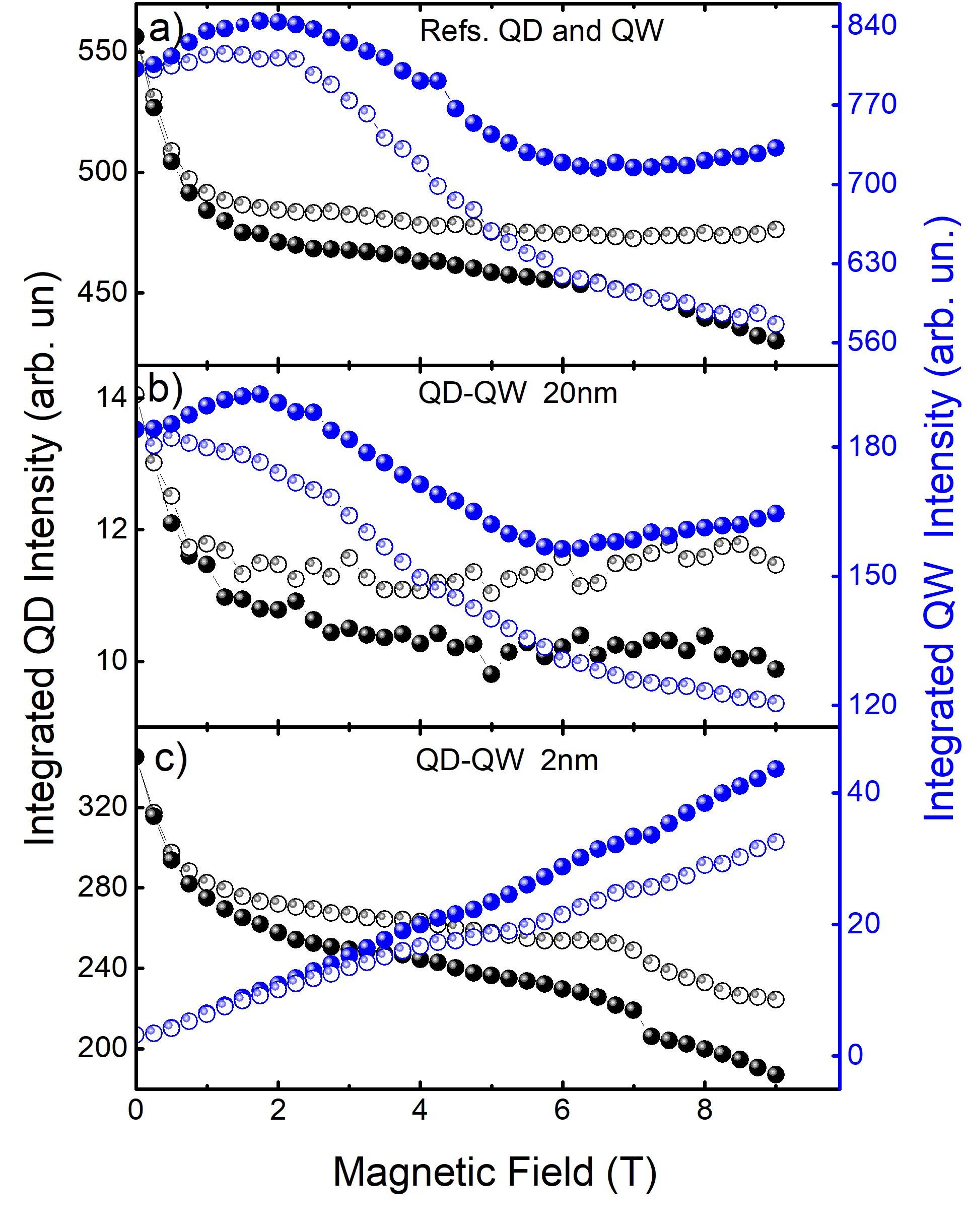}
\end{center}
\par
\vspace{-0.5cm} \caption{(a) Integrated PL intensities for QDs (left axes) and QW (right axes) ground state versus magnetic field measured in: (a) reference QD and QW, (b) QD-QW hybrid dot-well structure separation d$_{sp}$ = 20 nm and (c) QD-QW hybrid dot-well structure with d$_{sp}$ = 2 nm at P$_{ex}$= 90 W. Filled and hollow spheres represents the circular $\sigma^{+}$ and $\sigma^{-}$ polarized luminescence detection, respectively.} 
\label{bandstructure_el}
\end{figure}

	Using our knowledge of the behavior of the reference samples in magnetic field we investigated the hybrid dot-well structures where the QD and QW layers are separated by a GaAs barrier of different thicknesses. We expect here the PL properties of the constituent QD and QW layers be similar to their reference examples, and will use this to understand the effects of the hybridization. Fig. 1(c) shows the PL spectra of the dot-well sample with d$_{sp}$ = 20 nm  under magnetic field (B = 0 and 9 T) at the same high excitation power, under which the reference samples were measured (see Fig. 1(a) and 1(b)). The 20 nm  barrier is thick enough to assure weak direct dot-well coupling, such that the dot and well layers should be considered independent. Here we find the zero field PL spectra to be very similar in position and shape to those of the reference samples. Noticeably, though the integrated intensities are significantly lower for both the QD and QW layers. This implies that a thick GaAs barrier substantially attenuates the excitation on the hybrid structure. As a result we observe distinctly only the $E_{QD}^{1}$ excited state of the QDs and a significantly weaker emission from the $e_{1}-hh_{1}$ in the QW. In addition, the magnetic field behavior becomes less pronounced for this 20 nm barrier sample. Nevertheless the QD first excited level under 9 T is splitted in two components which can be separated by spectral deconvolution (Fig. 1(a)). The PL spectrum of the QW shows three LL emissions which implies in a high density of excitons in the QW Fig. 1(d). As a result of these observations we conclude that for the thick barrier the QD and QW structures respond very much like the reference samples.
			
	This changes dramatically for thin GaAs spacers ($d_{sp}$ = 2 nm). The hybrid structures with thinner barriers belong to the class of structures with strong direct coupling leading to a hybridization of the QD and QW excitonic states \cite{8, 9, 10, 11, 12}. It has been shown and can be estimated from Fig. 1 that the spectral ranges corresponding to the QD third excited state, and the QW ground state overlap each other \cite{10,12}. As a result, due to the thin spacer the carriers can resonantly tunnel from the QW to the QD through the overlapped states, relax to the QD ground state and emit in a spectral range that differs substantially from emission range of the QW. Figure 1(e) shows the PL spectra of the hybrid structure with d$_{sp}$ = 2 nm measured with P = 90 $\mu$W excitation without a magnetic field (blue lines) and at a magnetic field of B = 9 T in the $\sigma^+$ and $\sigma^-$ polarizations (black and red lines, respectively). In contrast to Fig. 1(c), the zero-field QD PL spectrum exhibits the pronounced structure of excited states up to $E_{QD}^{3}$ indicating strong optical pumping to the QD layer leading to more PL intensity than seen in the hybrid structures d$_{sp}$= 20nm. At the same time, however, the zero-field QW emission is comparatively weaker and it vanishes completely at low power. This behavior indicates that the QW to QD carrier transfer is very efficient and completely depletes the population of the QW states at low excitation levels. Now, if the magnetic field is applied at B = 9 T, the QD emission gradually decreases whereas the QW becomes high enough to develop even the $N = 1$ LL transition in the QW PL spectrum  (Fig. 1(f)).
			
	Figure 2(a) shows the integrated PL intensities for the QDs and QW reference samples as functions of the applied magnetic field where they are both seen to become generally weaker with increasing field. For the QW a slight increase in the integrated intensity at low fields can be seen, however as the magnetic field increases and the Landau levels appear, the FWHM (not shown here) starts to decrease and a drop in the integrated intensity is observed. Since the QD-QW system with d$_{sp}$= 20nm (Fig. 2(b)) exhibits weak coupling, the same behavior of the integrated intensities versus magnetic field is observed.
			
	The QDs and the QW integrated PL intensities as a function of the applied magnetic field for the hybrid structure with d$_{sp}$= 2nm are shown on Figure 2(c). When contrasted with the reference samples (Fig. 2(a)) the QD emission has a comparable decrease in its intensity.  On the other hand, for the QW emission, an intensity increase is observed, which is the opposite from what is depicted on the reference sample. We assume that such anticorrelated change of the integrated intensities indicates a very efficient magnetically controlled reduction of the dot-well coupling. This blocks the carrier transfer from the QW into the QD system, and thus enhances the QW PL by increasing the exciton density in the QW, while causing the QD PL to reduce by cutting off its exciton source.

			\begin{figure}[h!]
\linespread{0.5}
\par
\begin{center}
\includegraphics[scale=0.3]{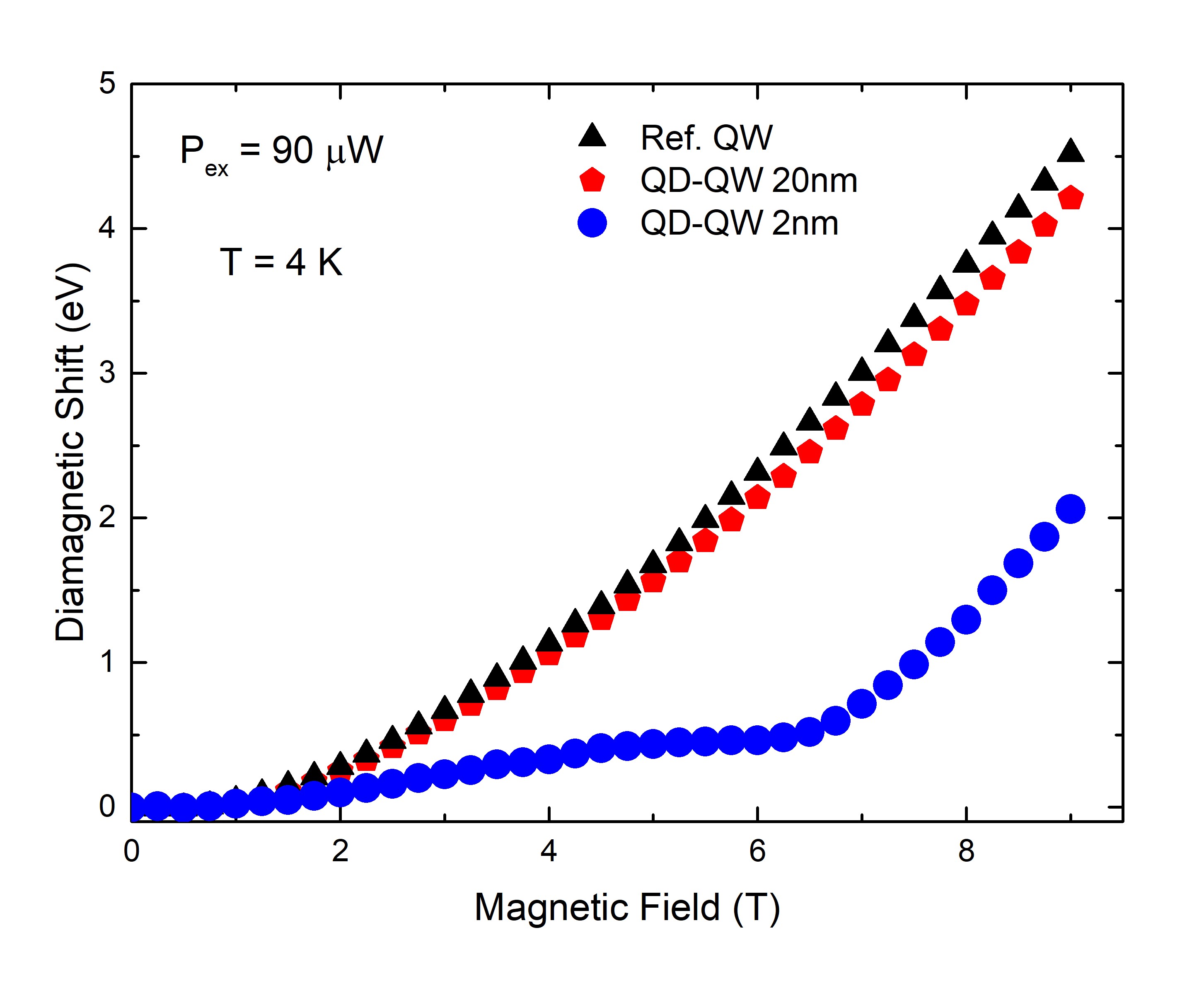}
\end{center}
\par
\vspace{-0.5cm} \caption{Diamagnetic shift versus magnetic field measured for the QW emission in the hybrid structures with d$_{sp}$= 2 nm, 20 nm and in the reference QW at P$_{ex}$= 90  $\mu$W.  }
\label{bandstructure_el}
\end{figure}
	
	The orbital Zeeman splitting of the QD states reaches only tenths of meV at a magnetic field of 9 T whereas the QW ground state shifts by 6 meV. We suggest that this difference in motion between the energy states of the 2D QW and those of the confined QDs with magnetic field results in a strong detuning of the 3$^{rd}$ QD excited state with the QW ground state. The tunneling then becomes non-resonant which significantly lengthens the time of carrier transfer between the QW and QD systems and weakens the dot-well coupling. Fig. 3 supports this conclusion. Here, the QW diamagnetic shift with magnetic field is shown for the reference QW and for the hybrid dot-well structures with different $d_{sp}$ values. It is known that strongly localized states exhibit smaller diamagnetic shifts than the less localized states\cite{23}. This is what we see in Fig. 3. The reference QW and the QW with $d_{sp}$ = 20nm both represent free 2D excitons with coinciding diamagnetic shifts. In the case of the d$_{sp}$ = 2 nm,  the diamagnetic shift reflects on the appearance of the dot-well coupling and reveals two ranges: the range of weakly varying shift at small magnetic fields indicating a regime of strong coupling and increased exciton localization; and the range of rapid growth at higher magnetic field indicating a regime of decreasing localization which destroys the resonant tunneling, and substantially decouples the dot-well structures.
			
	The observed decrease of the integrated PL intensity with magnetic field parallel to the growth direction has been observed before in arrays of InAs QDs in Ref.[18]. Here, it was established that the magnetic field in the Faraday geometry reduces the lateral transport to the dots since the field localizes carriers in wetting layer (WL) potential fluctuations with depths of a few meV. This formally means that the state population of the QDs at the same excitation power is lower with an applied magnetic field, while at the same time, the population of the WL states increases. As a result, the QD emission is reduced, whereas the WL PL is enhanced \cite{18}. We cannot completely exclude this mechanism in the quenching of our QD PL with magnetic field, however, at the best, it is not dominant in our case. Indeed, we do not observe any PL  enhancement in the range of the WL transition (1.45 eV). Moreover, the magnetic field can enhance the localization of carriers in the considerably deeper QD potential, as compared with WL potential fluctuations, thus favoring an increase of the QD PL intensity with increasing magnetic fields in Faraday geometry. Additionally the exciton lifetime reduces with an applied magnetic field, again contributing to the enhancement of the QD PL intensity.
			
	Enhancement of a QW PL with a magnetic field has been observed as well \cite{19} and attributed to the magnetic field induced compression of the wave function and corresponding increased oscillator strength. As a result the observed PL intensity was found to increase 1.5 times in an InGaAs/InP QW with an applied magnetic field of 7 T. A compression of the in-plane wave functions of the carriers in a magnetic field perpendicular to the QW was also revealed in the time decay measurements of QWs\cite{20}. This mechanism of QW PL enhancement cannot be ignored in our case, but it would not explain the 8 and 10-fold(red and black arrow fig. 1(f) ) increase of the PL amplitude observed in our hybrid structures.
			
	There are several possibilities for the physical mechanisms of the magnetically reduced coupling observed in our hybrid dot-well structures. They must include jointly the effects of strain, confinement, and magnetic field on the valence band which contributes to the magnetoexciton state in the hybrid structure. This mechanism must also reproduce the experimentally observed ratio of the WL PL over the integrated PL intensity of the QDs which increases nonlinearly with the magnetic field as well as the transport properties affecting the carrier capture by the QDs \cite{21}. Of particular importance is the representation of the highly excited states of the InAs QDs in a magnetic field. Indeed it has been demonstrated that the overall pattern of the magnetic field evolution of the emission lines related to these states resembles a single-particle Fock-Darwin diagram \cite{22}.
	
\section{Conclusions}
In this study we carried out a magnetoPL study of the carrier transfer in hybrid InAs/GaAs quantum dot(QD)-InGaAs quantum well (QW) structures with varying barrier thickness. The measurements were performed with the magnetic field parallel to the growth direction (the Faraday geometry) up to 9 T and low temperature (4 K). For excitation densities sufficiently high to observe the QW PL we found a strong dependence of the QW PL intensity on magnetic field both for weak and strong coupling between the dots and the well. The observed exchange of PL intensity from the QDs to the QW with magnetic field is attributed the breaking of their resonant coupling. This is the result of a change of in-plane carrier dynamics in both the QW and WL. Both the formation of Landau levels and the change of coupling between QD excited states and exciton states in the QW and the WL due their different diamagnetic shifts ultimately bringing them out of resonance.

\section{Acknowledgment}
The authors gratefully acknowledge the financial support of the following agencies: National Science Foundation of the U.S. (Grants No. DMR-1008107 and No. DMR-1108285),CSIC grant I-COOP-2017-COOPB20320  FAPESP (grants $\#$ 2013/18719-1, 2014/07375-2, 2014/19142-2, 2018/01914-0), CNPq(164765/2018-2) and CAPES (Finance Code 001).

\bibliographystyle{apsrev}

\bibliography{bibliography}

\end{document}